# Special relativity may account for the spacecraft flyby anomalies

Jean Paul Mbelek[(1)]

[(1)]Service d'Astrophysique, CEA-Saclay, Orme des Merisiers, 91191 Gif-sur-Yvette, France

March 15, 2009

Abstract

The empirical formula proposed by J. D. Anderson *et al.* [1] to reproduce the data on the Earth flyby anomalies is derived from special relativity (SR). The transverse Doppler shift together with the addition of velocities account for the Doppler data. Time dilation together with the addition of velocities account for the ranging data.

## 1 Introduction.

Galileo (in 1990 and 1992), NEAR (in 1998), Cassini (in 1999), Messenger (in 2005) and Rosetta (in 2005) spacecrafts showed an unexpected frequency change in the radio Doppler data during Earth flyby which are about ten times larger than the expected errors in measurement. Similar anomaly is found in the ranging data. The largest anomaly was recorded for NEAR (13.5 mm/s in excess) [1, 2]. During these Earth flyby, the incoming asymptotic velocity $\underline{V}_{\infty\,i}$ and outgoing asymptotic velocity $\underline{V}_{\infty\,o}$ of the spacecraft seemingly differed in magnitude not only by direction. The speeds $\underline{V}_{\infty\,i}$ and $\underline{V}_{\infty\,o}$ are respectively deduced by converting the measured two-ways Doppler shifts $(\delta\nu/\nu)_i$ and $(\delta\nu/\nu)_o$ into velocities from the usual nonrelativistic Doppler formula, thereby the so-called flyby anomaly, $\Delta V_\infty$. Recently, J. D. Anderson *et al.* [1] have proposed an empirical formula that accurately reproduces most of the Earth flyby anomalies observed as yet. This formula involves respectively the declination $\delta_i$ and $\delta_o$ of the spacecraft's incoming and outgoing asymptotic velocities with respect to the geocentric reference frame and reads

$$\Delta V_\infty / V_\infty = K\,(\cos\delta_i - \cos\delta_o), \qquad (1)$$

where $K = 2\omega_E R_E/c = 3.099 \times 10^{-6}$ and, $\omega_E = 7.292115 \times 10^{-5}$ rad/s and $R_E = 6371$ km denote respectively the angular velocity and the mean radius of the Earth. If so, then the rotation of the Earth should be included in the analysis of the Earth flyby anomalies.

However, no underlying physical explanation for the anomaly based on conventional phyics has been put forward thus far to support the latter formula. Current proposals rather assume new physics is the cause of the anomaly as for instance light speed anisotropy [3], a modification of inertia [4] or dark matter [5]. Since the Earth's angular velocity and radius enter in the formula (1), the Lens-Thirring effect predicted by general relativity (GR) could be an explanation but this turns out to be negligibly small. Also, Lämmerzahl *et al.* have ruled out many potential sources of the flyby anomalies [6]. Furthermore, let us emphasize that the software the Jet Propulsion Laboratory uses is relativistic but not fully relativistic, in particular as regards to the rotation of the Earth and the other planets because of practical difficulties in numerical calculations although the problems are well-understood theoretically (S. A. Klioner recalls that : <<Earth rotation is the only astronomical phenomenon which is observed with a high accuracy and which has no consistent relativistic model.>>) [7]. In the following, we show that SR transverse Doppler shift together with the addition of velocities are sufficient to explain the Doppler data of the flyby anomalies. Thus, GR does not need to be questioned and the flyby anomaly is merely due to an incomplete analysis using conventional physics. Moreover, we show that the SR time dilation together with the addition of velocities, on account of the Earth's rotation, yield relation (1) from ranging data.

## 2 Kinematics of the spacecraft flyby anomalies.

The photons of the communication signals between the ground based antennas and the spacecraft are subject to the transverse Doppler shift on account of the relative velocity, $\underline{\mathbf{W}}$, between the spinning Earth and the orbiting spacecraft. We argue that this well-known special relativistic effect has not been properly accounted for in this context. Let $\nu_0$ and $\nu$ be respectively the frequencies of the emitted and received photons. The ground based antennas of the Deep Space Network (DSN) station at latitude $\phi_S$ are endowed with the rotational velocity $V_E = \omega_E R_E \cos\phi_S$ with respect to the geocentric reference frame. Hence, the one-way transverse Doppler shift reads

$$\nu = \nu_0 (1 - W^2/c^2)^{1/2}, \qquad (2)$$

where $\underline{\mathbf{W}} \cong \underline{\mathbf{V}}_\infty + \underline{\mathbf{V}}_E$, according to the addition of velocities in the low velocity approximation. Hence, the two-ways transverse Doppler shift for a given speed of the spacecraft with respect to the geocentric reference frame reads

$$\delta\nu/\nu_0 = 2 [(1 - W^2/c^2)^{1/2} - 1] \cong - V_\infty^2/c^2 - V_E^2/c^2 - 2 \underline{\mathbf{V}}_\infty \cdot \underline{\mathbf{V}}_E/c^2 \qquad (3)$$

In case the declinations $\delta_i$ and $\delta_o$ of respectively the incoming (i) and outgoing (o) osculating asymptotic velocity vector differ with respect to Earth's equator, relation (3) above yields an overall Doppler shift in excess given by

$$\Delta\nu/\nu_0 = (\delta\nu/\nu_0)_o - (\delta\nu/\nu_0)_i = 2 (\underline{\mathbf{V}}_{\infty\,i} - \underline{\mathbf{V}}_{\infty\,o}) \cdot \underline{\mathbf{V}}_E/c^2, \qquad (4)$$

by setting $V_{\infty\,o} = V_{\infty\,i} = V_\infty$, which means that there is no real flyby anomaly.

Now, the SR transverse Doppler shift (4) may be cast into a nonrelativistic Doppler shift-like relation, namely $\Delta\nu/\nu_0 = \Delta V_\infty \cos\alpha/c$, where $\alpha$ would represent the angle made by the direction from the spacecraft to a DSN station with the difference vector $\Delta\underline{\mathbf{V}}_\infty = \underline{\mathbf{V}}_{\infty\,o} - \underline{\mathbf{V}}_{\infty\,i}$. Arranging relation (4), one obtains :

$$\Delta V_\infty / V_\infty = K (\cos\phi_S / \cos\alpha)(\cos\delta_i - \cos\delta_o). \tag{5}$$

Clearly, relation (5) reduces to (1) for flybys such that $\alpha \cong \phi_S$.

As regards the ranging data, let us recall that time delay is the observable of the ranging data and any time delay in excess, $\Delta T$, may be interpreted as an increase in velocity (see M. T. Jaekel and S. Reynaud [8]), $\Delta V_\infty$, with respect to a DNS station given by

$$\Delta V_\infty \cos\alpha = c \, d(\Delta T)/dt. \tag{7}$$

Also, as one knows, any change in velocity implies a change in position but not necessarily a change in the magnitude of the velocity. Hence, the observation of the so-called flyby anomalies is still expected in ranging data even assuming $V_{\infty o} = V_{\infty i}$. Now, SR time dilation yields a two-ways time delay in excess, $\Delta T$, given by

$$d(\Delta T)/dt = 2 \left[ (1 - \underline{W}_{\infty i}^2/c^2)^{-1/2} - (1 - \underline{W}_{\infty o}^2/c^2)^{-1/2} \right]. \tag{8}$$

Expanding relation (8) to the first order in $\underline{W}_{\infty i} \cong \underline{V}_{\infty i} + \underline{V}_E$ and $\underline{W}_{\infty o} \cong \underline{V}_{\infty o} + \underline{V}_E$, one gets

$$d(\Delta T)/dt = 2 (\underline{V}_{\infty i} - \underline{V}_{\infty o}) \bullet \underline{V}_E / c^2, \tag{9}$$

by setting $V_{\infty o} = V_{\infty i} = V_\infty$ (still there is no real flyby anomaly). Combining relations (7) and (9) yields relation (5).

## 3 Conclusion.

Reconsidering the analysis of the Earth flyby anomalies observed up to now, and for which J. D. Anderson *et al.* have proposed an empirical formula (1), we have found that the SR time dilation, the SR transverse Doppler shift and the addition of velocities (to account for the Earth's rotation) all together provide the correct solution. Hence, we conclude that spacecraft flybys of heavenly bodies may be viewed as a new test of SR which has proven to be successful near the Earth. We call for a further follow-up of the spacecraft trajectory to test relation (5) well beyond the radius of closest approach.